\documentclass[%
 reprint,
 amsmath,amssymb,
 aps,
]{revtex4-2}

\usepackage{graphicx}%
\usepackage{dcolumn}%
\usepackage{bm}%
\usepackage{xcolor}
\usepackage{ulem}

\begin{document}

\preprint{APS/123-QED}

\title{Does nematic order allow groups of elongated cells to sense electric fields better?}%

\author{Kurmanbek Kaiyrbekov}
\author{Brian A. Camley}%
\affiliation{%
 Department of Physics \& Astronomy, Johns Hopkins University, Baltimore, Maryland 21218, USA
}%

\begin{abstract}
Collective response to external directional cues like electric fields plays a pivotal role in processes such as tissue development, regeneration, and wound healing. In this study we focus on the impact of anisotropy in cell shape and local cell alignment on the collective response to electric fields. We model elongated cells that have a different accuracy sensing the field depending on their orientation with respect to the field. Elongated cells often line up with their long axes in the same direction – ``nematic order'' – does this help the group of cells sense the field more accurately? We use simulations of a simple model to show that if cells orient themselves perpendicular to their average velocity, alignment of a cell’s long axis to its nearest neighbors’ orientation can enhance the directional response to electric fields. However, for cells to benefit from aligning, their accuracy of sensing must be strongly dependent on cell orientation. We also show that cell-cell adhesion modulates the accuracy of cells in the group.
\end{abstract}

\maketitle

\section{\label{sec:Intro}Introduction}
Electric fields are ubiquitous in the extracellular environments of plants, animals, and humans. They play crucial roles in biological processes such as development, physiology, regeneration, and pathology \cite{mccaig2009electricalDimCellScience, LEVIN2007261}. Cells, both individually and collectively, follow electrical fields; this response is called galvanotaxis or electrotaxis \cite{zajdel2020scheepdog, shim2021overriding}. The exact extent and manner of collective cell response to field hinge on factors like cell type, cluster size, electric field strength, and intercellular interactions \cite{zajdel2020scheepdog, li2012cadherin, lalli2015collectiveMigGreaterSensitivity}. 

Galvanotaxing cells tend to align their long axes perpendicularly to the field \cite{zajdel2020scheepdog, HAMMERICK201012_Alignment, Lang_2021_DCFieldalign, Bunn2019DCEFPerpAlignment, GUO20102320EFieldFirbroblasts, wolf2022short-term-bioelstrim}. We hypothesize that this may allow cells to better sense the electric field -- i.e. that a cell's accuracy at sensing the field direction is better if the cell is elongated perpendicular to the field. Theoretical studies on chemical gradient sensing have demonstrated that elliptical cells possess higher accuracy in this orientation \cite{GeometryBiasGradientSensing}, and preliminary results extending our model of \cite{nwogbaga2023physical} to galvanotaxis also suggest galvanotaxing cells may -- in some circumstances -- be better sensors in this orientation. If elongated galvanotaxing cells are more accurate when perpendicular to the field, how would this affect a group of cells' ability to sense an electric field?
This question is particularly interesting because even in the absence of external cues like electric fields, confluent elongated cells tend to align with one another, resulting in the emergence of local nematic order \cite{duclos2014perfectNemOrdInSpindleShapedFibroblasts, nematicOrderInConfinedPopOfFibroblasts,GuillamatIntegerMyoblast,kaiyrbekov2023migration,saw2017celldeathandextrusion}. %
If elongated cells are better sensors when they are perpendicular to the field, nematic cell-cell alignment may let groups of cells cooperate to ensure they are aligned in the best direction, improving collective galvanotaxis. 

We develop a minimalist model of groups of elongated cells with anisotropic precision in sensing electric fields. We find that when the precision of field direction estimation strongly depends on cell orientation, such that perpendicular alignment to electric fields makes cells' estimates more accurate, cells significantly improve at following applied field by aligning perpendicular to their own averaged velocity (Section \ref{sec:results}A-B). Additionally, we find that local nematic interactions between cells can in some circumstances increase cluster directionality (Section \ref{sec:results}C). Finally, we show that strong cell-cell adhesion enhances directional response of cells to electric fields and we present mechanisms that could potentially disrupt this response, suggesting potential explanations for conflicting results on the necessity of cell-cell adhesions for collective galvanotaxis (Section \ref{sec:results}D). We have chosen a deliberately minimal model, and expect many of our results may also be informative about how elongated cells can cooperate to sense different types of gradients \cite{camley2018collective}. 
\begin{figure*}
\centering
\includegraphics[width=0.95\textwidth]{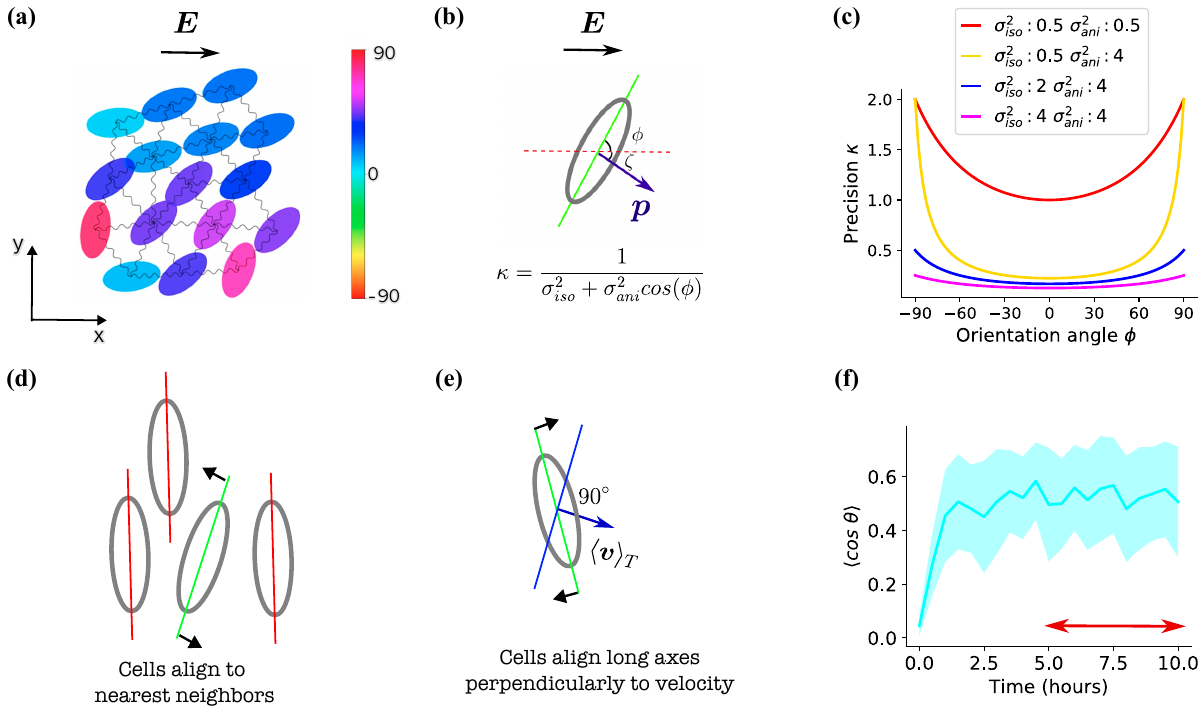}%
\caption{\label{fig:intro_fig} Description of cell parameters and simulation model. \textbf{(a)} The depicted snapshot illustrates a simulation box, showcasing 16 interacting cells interconnected by springs. The electric field within this model is oriented along the positive x-axis. Each cell is uniquely color-coded to represent the angle formed between the x-axis and the cell's major axis. \textbf{(b)} A cell exposed to field $\mathbf{E}$ estimates the direction of the field and polarizes towards its estimate. A cell's precision $\kappa$ in detecting the electric field direction, depends on its orientation angle $\phi$, where $\phi$ is the shortest angle between the cell's longitudinal axis (green line) and the x-axis (red dashed line). With this definition, $\phi$ will be between $-90^\circ$ and $+90^\circ$. \textbf{(c)} Cell precision $\kappa$ as a function of cell orientation for different values of intrinsic ($\sigma_\textrm{iso}^2$) and anisotropic ($\sigma_\textrm{ani}^2$) variances. Cells oriented horizontally ($\phi = 0$) have the lowest precision and cells with vertical orientation ($\phi = 90^\circ$) have the highest precision. \textbf{(d)} A diagram illustrating the tendency of a cell, with its long axis represented by a green line, to align with neighboring cells, whose long axes are depicted with red lines. \textbf{(e)} A diagram depicting the tendency of a cell to align its orientation, illustrated by a green line, perpendicularly to the cell's average velocity. The orientation orthogonal to the velocity is represented by a blue line. \textbf{(f)} The graph presents the temporal evolution of the average directionality of cell groups across 40 simulations, with shaded areas representing the standard deviations calculated over these simulations. Each simulation featured 64 cells. The red arrow spans the final 5 hours and we report the mean over this period as the final steady state directionality. }
\end{figure*}

\section{Model}
 We use a two-dimensional self-propelled particle model to describe cell behavior in the presence of an electric field ${\bf E} = E \hat{x}$ (Fig. \ref{fig:intro_fig}a). We model {elongated cells as particles characterized by positions $ {\bf r}^i = (x^i, y^i)$ and the orientation of the long axis $\phi^i$ for the $i^\textrm{th}$ cell (Fig. \ref{fig:intro_fig}b).} Each cell will also have a polarity direction $\zeta^i$ -- the direction the cell is propelling itself in. Cells exhibit an orientation-dependent accuracy in sensing the field, are interconnected through spring-like adhesions, align themselves perpendicular to their velocity, and orient their long axes in alignment with those of their neighbors (Fig. \ref{fig:intro_fig}a-e).

\subsection{Translational motion}
 Cells have an internal biochemical polarity that sets the direction of their motion. This polarity arises due to asymmetry between the back of the cell where myosin contractions pull the rear, and front where filopodia and/or lamellipodia extend the cell frontier \cite{ananthakrishnan2007forcesBehindCellMovement}. The polarity of cell $i$ ${\bf p}^i = (p_x^i, p_y^i) = p_0(\cos \zeta^i,\sin \zeta^i)$ is the velocity that the cell would have in the absence of interactions with other cells. The polarity vector has a magnitude of {$p_0 = 1 \ \mu m / min$}  (equivalent to $60 \ \mu m / h$), roughly consistent with typical single-cell speeds \cite{tschumperlin2013fibroblastsGroundTheyWalkOn, zarkoob2015substrateDepKeratinocyteSpeed}. We evolve cell positions according to over-damped Newtonian  equations of motion:
\begin{equation}
\frac{\partial {\bf r}^i}{\partial t} = {\bf p}^i + \sum_{j \underset{n}{\sim} i} {\bf F}^{ij}
\label{eq:eqofmotion}
\end{equation}
where ${\bf F}^{ij}$ is an inter-cellular force (e.g. adhesion and volume exclusion) by cell $j$ on cell $i$, and the sum is over the neighbors  that are within the interaction cutoff distance $r_{c}$ from the cell $i$ ({the interacting neighbors $j$ of cell $i$ are denoted as $j \underset{n}{\sim} i$}). If the distance between two interacting cells exceeds the interaction cutoff distance $r_c$ they cease to interact -- modeling a disconnection of cell-cell adhesion. The force arises due to neighboring cells interacting with a harmonic spring:
\begin{equation}
{U}^{ij} = \frac{1}{2} k (  r^{ij}  - r_{eq})^2
\label{eq:harmpotential}
\end{equation}
where ${r^{ij}} = \lvert {\bf r}^i - {\bf r}^j \rvert$ is the distance between cells $i$ and $j$, and $r_{eq}$ is a parameter that sets an equilibrium distance between cells. The force exerted by cell $j$ on cell $i$ is given by
${\bf F}^{ij} = - \nabla_i {U}^{ij}$. Here, the spring constant k, which we can think of as setting the cell-cell adhesion strength is expressed in units of 1/time -- implicitly this means that we have absorbed a friction coefficient into the value of k instead of writing it in Eq. \ref{eq:eqofmotion}. We note that the spring interaction does not depend on the particle orientation -- the effect of the orientation is solely on the the cell's ability to sense the field direction and on cell-cell orientational alignment.

\subsection{Cell polarity and field sensing}

We assume that cells estimate the field direction and then polarize in the direction of their estimate. (This neglects any dynamics of reaching this estimated direction \cite{nwogbaga2023coupling,prescott2021quantifying}.) We also assume that precision of a cell's estimate depends on its orientation such that it has highest precision when oriented orthogonally to electric field, and lowest precision if its long axis is parallel to the field. We model this estimation process as the cell drawing its estimated direction $\zeta^i$ from a von Mises distribution centered around the true electric field angle with respect to x-axis (which is 0), with a width controlled by the precision $\kappa^i$ which depends on cell $i$'s orientation, i.e. $\zeta^i \sim VM(0, \kappa^i)$. The von Mises distribution is a generalization of the normal distribution to orientations \cite{mardia2009directional}, and has a form $p(\zeta) \sim e^{\kappa \cos(\zeta)}$ -- i.e. in the limit of large $\kappa$ it corresponds to a normal with variance $\sigma^2 = 1/\kappa$ -- and we will often refer to errors in terms of $\sigma^2$. 
The cell polarizes in the direction of its estimate $\zeta^i$:
\begin{equation}
{\bf p}^i  = \cos (\zeta^i) \hat{x}  +  \sin( \zeta^i) \hat{y} %
\end{equation}
 (see Fig. \ref{fig:intro_fig}b).

 The precision of the estimate $\kappa_i$ is orientation-dependent. We hypothesize that cells exhibit the highest precision when perpendicular to the electric field ${\bf E} = E \hat{x}$ (vertical orientation) and lowest precision in parallel (horizontal) orientation:
 \begin{equation}
     \kappa^i = \frac{1}{\sigma_\textrm{iso}^2 + \sigma_\textrm{ani}^2 \cos\phi^i}
 \end{equation}
  where the cell orientation $\phi^i \in [-\pi/2, \pi/2]$ is the smallest angle between the cell's long axis and the x-axis (Fig. \ref{fig:intro_fig}b). Hence, a cell $i$ in vertical orientation ($\phi^i = \pm \pi/2$) would have $\kappa^i = 1/\sigma_\textrm{iso}^2$ -- or an angular error of $\sigma_\textrm{iso}^2$ -- while in horizontal alignment ($\phi^i = 0$) would  have a precision of $\kappa^i = 1/ (\sigma_\textrm{iso}^2 + \sigma_\textrm{ani}^2)$ -- or an angular error of $\sigma_\textrm{iso}^2 + \sigma_\textrm{ani}^2$. %
  The $\sigma_\textrm{iso}^2$ is the orientation-independent intrinsic variance -- the baseline accuracy in estimation -- and  $\sigma_\textrm{ani}^2$ characterizes the additional variance stemming from the anisotropy of cell shape.  The orientation dependence of precision is most pronounced when the difference between values of $\sigma_\textrm{iso}$ and $\sigma_\textrm{ani}$ is substantial (see Fig. \ref{fig:intro_fig}c).

Cells can make an estimation of the electric field multiple times over our simulation. How often should we expect this estimation to happen? Galvanotaxis -- in our best current understanding -- requires redistribution of proteins or other sensor molecules on the cell membrane \cite{allen2013electrophoresis,nwogbaga2023physical}. Membrane proteins on a cell of radius $R$ redistribute via diffusion in a characteristic time $\tau_{\text{forget}} = R^2 / D$, where $D$ is the protein diffusion constant. Averaging time-correlated estimates of the electric field direction for a duration less than $\tau_{\text{forget}}$ does not improve accuracy \cite{nwogbaga2023physical}. Therefore, we assume that cells estimate field direction every $\tau_{\text{forget}}$ minutes and keep the polarity to be constant until a new estimate is made. We estimate $\tau_\textrm{forget}$ to be on the order of 10 minutes \cite{nwogbaga2023physical,allen2013electrophoresis}.

\begin{figure*}
\centering
\includegraphics[width=0.95\textwidth]{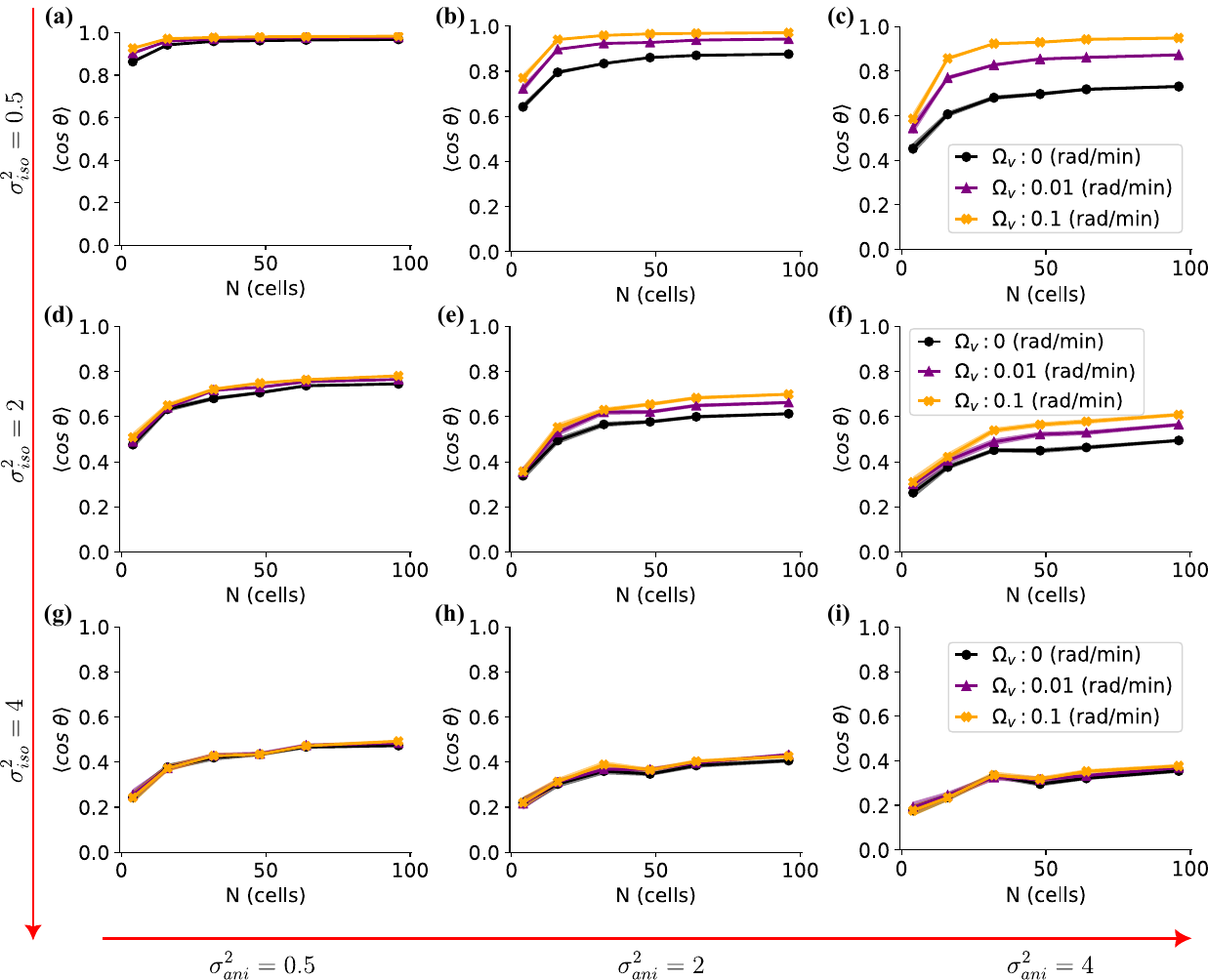}%
\caption{\label{fig:variance_dep}The average directionality $  \langle \cos\theta \rangle$ for various combinations of isotropic $\sigma_\textrm{iso}^2$ and anisotropic $\sigma_\textrm{ani}^2$ variances and alignment rates to velocity.  
Isotropic component ($\sigma_\textrm{iso}^2$) changes across  rows (top  to bottom) and anisotropic component ($\sigma_\textrm{ani}^2$) across rows (left to  right) with  specific values shown at right side and bottom of the figure (i.e. {the panel} (f) show directionalities for $\sigma_\textrm{iso}^2 = 2, \sigma_\textrm{ani}^2 = 4$ ). The averages are over 40  simulations at the interaction strength of $k$ = 0.2 $\textrm{min}^{-1}$. For each simulation the reported directionality is the steady state average over the final 5 hours  of simulation as shown in Fig. \ref{fig:intro_fig}f. Results for different vaues of alignment rates to average velocity are color coded $\Omega_v = 0$ rad/min (black), $\Omega_v = 0.01$ rad/min (purple), and $\Omega_v = 0.1$ rad/min (orange). The averaging time $T$ for velocity is set to 1h.  The shaded areas represent standard errors, although they may not be easily discernible due to their small size. }
\end{figure*}
\subsection{Cell reorientation}

Elongated cells in confluence exhibit a tendency to align with neighboring cells, resulting in the formation of a local nematic order \cite{maroudas2021HydraMorpho, GuillamatIntegerMyoblast, kaiyrbekov2023migration, saw2017celldeathandextrusion} (see Fig. \ref{fig:intro_fig}d). Additionally, when exposed to an electric field, many cell types, even in isolation, will follow the field but orient themselves orthogonally to it \cite{zajdel2020scheepdog, Lang_2021_DCFieldalign, Bunn2019DCEFPerpAlignment, GUO20102320EFieldFirbroblasts, wolf2022short-term-bioelstrim}. In other words, on average, cells align their long axis perpendicularly to their velocity vector (depicted in Fig. \ref{fig:intro_fig}e). We incorporate these tendencies into the equations of motion for $\phi^i$, the orientation of the long axis of cell $i$:
\begin{eqnarray}
\frac{\partial {\phi^i}}{\partial t} = &&- \ \Omega_{v} \sin(2[\phi^i - (\alpha_{\langle \boldsymbol{v}^i \rangle_T} + \pi/2)])\nonumber\\&& - \ \Omega_n \sum_{j \underset{n}{\sim} i} \sin(2[\phi^i - \phi^j]) + \sqrt{2 D_r} \xi_r(t)
\label{eq:rot_eqofmotion}
\end{eqnarray}
The first term in Eq. \ref{eq:rot_eqofmotion} represents the cell's orientation tending to rotate to be $\pi/2$ away from the direction of the time-averaged velocity $\alpha_{\langle \boldsymbol{v}^i \rangle_T}$, with a rate of alignment $ \Omega_{v} $. Here, we define $\alpha_{\langle \boldsymbol{v}^i \rangle_T}$ as the angle between $\langle \boldsymbol{v}^i \rangle_T$ and horizontal axis, where $\langle \boldsymbol{v}^i \rangle_T$ is the velocity averaged over past period $T$, i.e. if $\partial \mathbf{r}^i/\partial t$ is the instantaneous velocity of the cell $i$ then $\langle \boldsymbol{v}^i \rangle_T = (1/T) \int_{t-T}^t (\partial \mathbf{r}^i/\partial t) dt $. 
Choosing orientation to tend to become perpendicular to velocity is a standard modeling choice for many elongated cells (see, e.g. \cite{nwogbaga2023coupling}). Here, we explicitly model the cell tending to become aligned to the time-average of velocity. This is sensible if cell velocity leads to accumulation of some internal polarity over a timescale $T$ (e.g. as by \cite{maiuri2015actin}). This would be equivalent to the cell's orientation being set by something controlled by the net displacement of the cell, as previously applied in collective migration by, e.g. \cite{kabla2012collective}.

The second term in Eq. \ref{eq:rot_eqofmotion} nematically aligns a cell's long axis with that of its neighbors $j$ with a rate of alignment $\Omega_n$ -- i.e. it ensures that cells' long axes are either parallel or antiparallel. Finally, $D_r$ is the rotational diffusion coefficient which describes spontaneous reorientation of the cell, and $\xi_r$ is a Langevin noise term with Gaussian probability distribution that has a zero mean $\langle \xi_r(t) \rangle = 0$ and no time correlation $\langle \xi_r(t_1)\xi_r(t_2) \rangle = \delta (t_1-t_2)$.

\section{Results\label{sec:results}}

The key parameters that we will measure in our simulations are cell directionalities -- the average cosine of the angle between the cell velocity and the field $\langle \cos \theta \rangle$ -- and cell speeds $\langle v \rangle$.  To be consistent with experiments which measure cell velocities over a fixed time window \cite{zajdel2020scheepdog, shim2021overriding, wolf2022short-term-bioelstrim, li2012cadherin, lalli2015collectiveMigGreaterSensitivity}, we also compute individual cell velocities at intervals of $30$ minutes. For instance, if the displacement of cell $i$ is $\Delta \mathbf{r}^i = \Delta x^i \hat{x} + \Delta y^i \hat{y}$ during 30 minutes, then the cell velocity is $\Delta \mathbf{r}^i/30 \textrm{min} = \langle \boldsymbol{v}^i \rangle_{T=30\textrm{min}} =  \lvert \langle \boldsymbol{v}^i \rangle_{T=\textrm{min}} \rvert ( \cos \theta^i \hat{x} + \sin \theta^i \hat{y})$, where $\lvert \langle \boldsymbol{v}^i \rangle_{T=30\textrm{min}} \rvert$ is the magnitude of the measured velocity.

We conduct $ \mathcal{S} = 40$ simulations of 10 hours for each parameter set. Initially, cells begin at random orientations -- so they cannot benefit from the increased accuracy when pointed perpendicular to the field. As the simulation evolves, cells estimate the direction of the electric field, update their polarities accordingly, and migrate toward this estimated direction. This movement is further influenced by the forces exerted by neighboring cells. Simultaneously, the cells adjust their orientations by aligning themselves perpendicularly to their averaged velocity and to nearest neighbors. Then, they update their estimates of field direction, and this cycle continues (Movie 1).  Measuring the directionality averaged across cells and across simulations, we observe that directionality increases over the initial 3-4 hours and then subsequently reaches a steady state (Fig. \ref{fig:intro_fig}f). During this time, cells become more ``vertically'' aligned. Through the rest of the paper, we will report the velocity and directionality averaged over the final 5 hours of simulation to characterize their steady-state values (red arrow in Fig. \ref{fig:intro_fig}f).

\subsection{Sufficient anisotropy in sensing accuracy and alignment to average velocity is necessary to see benefits of favorable alignment}

We hypothesized that cells are more accurate sensors when perpendicular to the field (``vertical''). To what extent does this assumption enhance cells' directional motion in a cluster? There are two key ingredients for sensing anisotropy to improve cluster accuracy: 1) the difference between sensing in the vertical orientation and the horizontal orientation must be large, and 2) cells must manage to reach the vertical orientation. Within our model, the term that drives cells toward the vertical orientation is the alignment perpendicular to velocity, controlled by $\Omega_v$. We modulate the rate of alignment to the average velocity ($\Omega_v$) across varying cell cluster sizes ($N$) for varying $\sigma_\textrm{iso}$ and $\sigma_\textrm{ani}$ in Fig. \ref{fig:variance_dep}. We initially set the alignment rate to nearest neighbors ($\Omega_n$) to zero to isolate the effect of $\Omega_v$, and will study nematic alignment effects later. From Fig. \ref{fig:variance_dep}, we see initially that an increase in group size ($N$) generally correlates with an improvement in directionality, regardless of the combinations of variances and alignment rates. 

In Fig. \ref{fig:variance_dep}, we observe that the alignment rate to velocity $\Omega_v$ -- which controls to what extent the cell has a vertical orientation -- only has a relevant effect in the upper-right four graphs, which are the cases where isotropic error $\sigma_\textrm{iso}$ is relatively low and anisotropic $\sigma_\textrm{ani}$ error is relatively high.  This is what we would expect. If the anisotropic variance is large while the isotropic error small, cells experience reduced precision if they significantly deviate from the favorable vertical orientation. Consequently, the advantages of faster alignment to average velocity become pronounced (see Fig. \ref{fig:variance_dep}b-c). By contrast, if cells have near-perfect estimates of the field direction independent of orientation (e.g. panel a, $\sigma_\textrm{iso}^2 = 0.5$ and $\sigma_\textrm{ani}^2=0.5$), then the cells within groups have near-perfect directionality independent of rates of alignment. Similarly, if cells' error is dominated by the isotropic component, e.g. $\sigma_\textrm{iso}^2 = 2$, cells experience a reduction in precision for all orientations, leading to an overall low level of directionality and no strong dependence on $\Omega_v$(Fig. \ref{fig:variance_dep}d-f). %

For the remainder of this paper, we adopt $\sigma_\textrm{iso}^2 = 2$ and $\sigma_\textrm{ani}^2 = 4$ as our default parameters -- panel f of Fig. \ref{fig:variance_dep}, Movie 1. We chose these values as leading to a trend of directionality with number of cells is roughly consistent with that found by \cite{li2012cadherin} for MDCK cells. The ratio of variance between the horizontal ($\sigma^2_\textrm{iso}+\sigma^2_\textrm{ani}$) and vertical  ($\sigma_{\textrm{iso}}^2$) orientations, which is $\frac{4+2}{2} = 3$ are roughly consistent with plausible numbers motivated by studies of chemotaxis \cite{GeometryBiasGradientSensing}, where the ratio of variance between the horizontal direction and vertical direction is $\approx 4$ for a cell with aspect ratio 2. 

Why does directionality improve with increasing $N$? Initially, this may not be surprising -- previous models of collective gradient sensing often found that the directionality of the cluster center of mass increases with cell number \cite{camley2018collective}. The cluster center of mass will almost always have a decreased noise as the number of cells increases simply from the law of large numbers because the center of mass motion reflects the average of many noisy motions of individual cells (see, e.g. \cite{camley2016emergent}; we repeat the basic argument in Appendix \ref{app:cm}). However, the directionalities we plot in Fig. \ref{fig:variance_dep} are for {\it single} cells -- not the center of mass. Why should directionality increase here? We argue this arises from cell-cell adhesion in our model. Even though the motion of the center of mass is independent of adhesion strength $k$, the spring strength $k$ controls the degree to which an individual cell follows the center of mass within a given time. In the limit of very large $k$, the cell cluster is essentially a rigid body -- each cell perfectly follows the center of mass. On the other hand, for tiny values of $k$, a cell acts as individual unit and weakly follows the center of mass. Hence, we expect minimal benefits from increasing cell cluster size in the case of weak adhesion but much stronger $N$ dependence in the case of strong adhesion. We repeat the simulations of Fig. \ref{fig:variance_dep} with very weak ($k$ = 0.05 $\textrm{min}^{-1}$) and strong ($k$ = 1 $\textrm{min}^{-1}$) adhesion strengths. Consistent with our expectations, directedness improves as cell groups get larger for strong adhesive forces (Fig. \ref{fig:variance_dep_k1}, Movie 2) while there are little benefits of increasing cell numbers for weakly interacting cells (Fig. \ref{fig:variance_dep_k0.05}, Movie 3). The essential role of cell-cell adhesion to gain a benefit in group galvanotaxis is consistent with experimental measurements showing E-cadherin is necessary for MDCK group galvanotaxis \cite{li2012cadherin}. We will explore the role of adhesion strength in more detail in Section \ref{sec:adhesion} below.

\subsection{Increasing velocity averaging time improves vertical alignment of cells and directionality}

\begin{figure}
\centering
\includegraphics[width=0.48\textwidth]{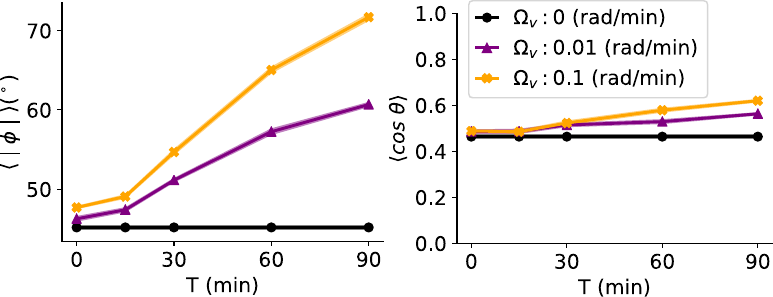}%
\caption{\label{fig:mem_dep} Cell orientations and directionalities as a function of velocity averaging time. In the left column, the absolute value of the cell orientation angle is presented, while the right column displays the directionality. The reported values represent averages across 40 simulations of 64 cells conducted with an interaction strength of $k$ = 0.2 $\textrm{min}^{-1}$ and $\sigma_\textrm{iso}^2 = 2 \ \& \ \sigma_\textrm{ani}^2 = 4$. The shaded areas represent standard errors of the mean.}
\end{figure}

Because cells align perpendicular to their average velocity, and cells are better sensors when perpendicular to the field, there is a natural feedback between orientation and velocity. As cells follow the applied field more accurately, their average velocity becomes more aligned to the electric field. This, in turn, means the cells become increasingly orthogonal to the field, improving their ability to sense the field. %
Because the instantaneous velocity of a cell will reflect both its self-propulsion and forces from neighboring cells, it will fluctuate around the true direction of the field. We expect that the velocity averaged over some time $T$ will thus have less variability and more reliably point toward the field-- so we expect that increasing the averaging time will increase the cell directionality as well as making the cells increasingly vertically-oriented. 

Running simulations while varying the averaging time $T$, we do see that cells become increasingly vertical and directional as $T$ increases (Fig. \ref{fig:mem_dep}). This dependence on averaging time, naturally, only happens if $\Omega_v > 0$, i.e. that cells orient perpendicular to their averaged velocity. The simulations of Fig. \ref{fig:mem_dep} are with our default adhesion level of $k$ = 0.2 $\textrm{min}^{-1}$. Cells connected with stiffer springs ($k$ = 1 $\textrm{min}^{-1}$) will have faster dynamics of cell-cell interactions, and we see pronounced improvement in alignment and directionality even for smaller averaging times (see Fig. \ref{fig:mem_dep_supp}b). Weakly interacting cells move more like individual units and while increasing averaging times improves alignment a little bit it does not relevantly improve the directedness of cells (Fig. \ref{fig:mem_dep_supp}a). This suggests that the relevance of the averaging time is to integrate over fluctuations of cell position due to relative motion from one cell to another, so that the averaged velocity better reflects the cluster center of mass motion.

\begin{figure*}
\centering
\includegraphics[width=0.95\textwidth]{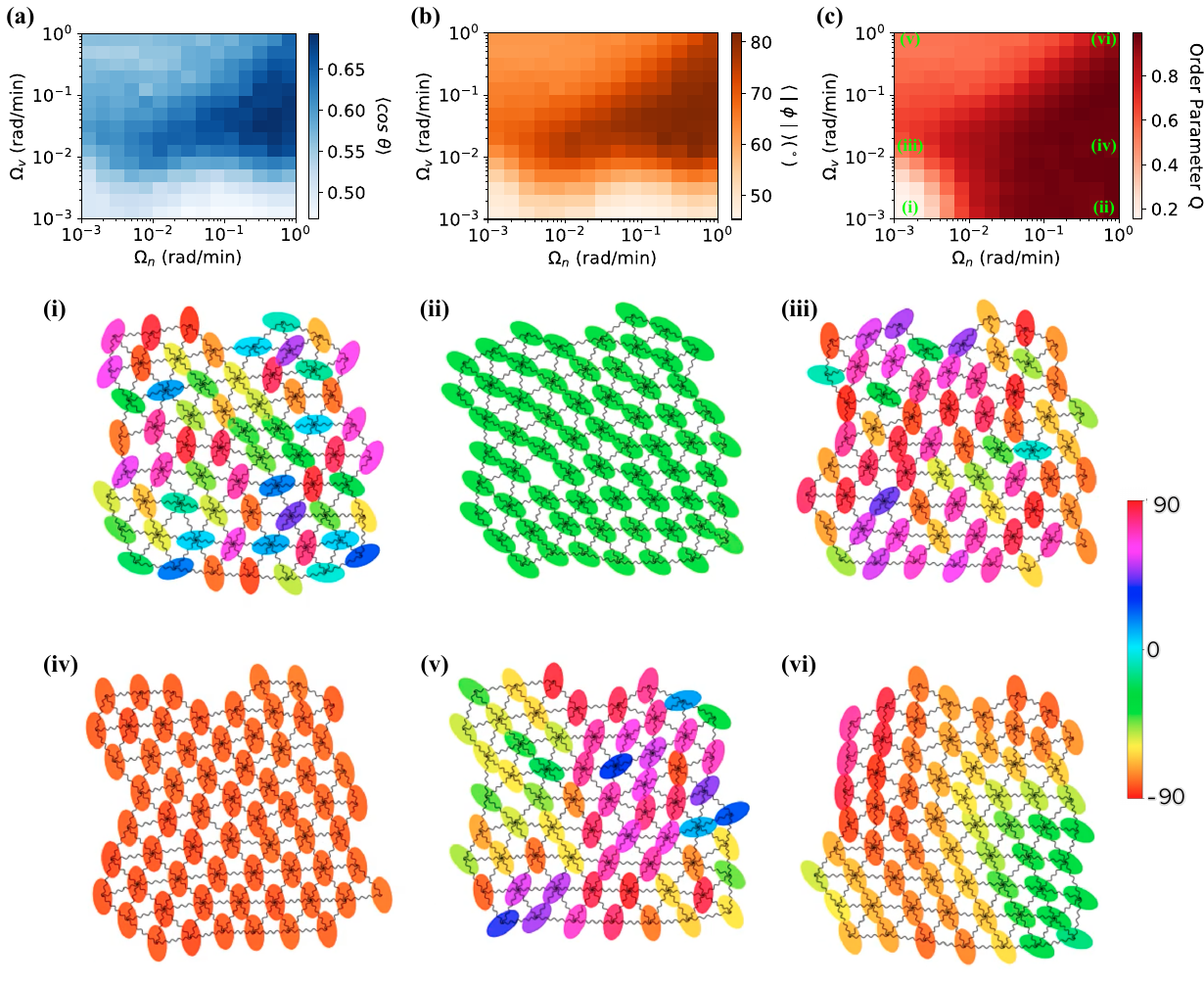}%
\caption{\label{fig:phase_diagrams} Effect of rates of alignment to velocity ($\Omega_v$) and neighbors ($\Omega_n$) on directionality \textbf{(a)}, absolute value of orientation angle \textbf{(b)} and order parameter \textbf{(c)}. Each grid value represents an average result over 40 simulations conducted with 64 cells at the interaction strength of k = 0.2 $\textrm{min}^{-1}$ with an averaging time 
$T$ = 1 h with colorbars indicating corresponding numeric values. Example simulation snapshots for alignment rate tuples of \textbf{(i)}, $\Omega_v = 0.001$ rad/min, $\Omega_n = 0.001$ rad/min; \textbf{(ii)}, $\Omega_v = 0.001$ rad/min, $\Omega_n = 1$ rad/min; \textbf{(iii)}, $\Omega_v = 0.012$ rad/min, $\Omega_n = 0.001$ rad/min; \textbf{(iv)}, $\Omega_v = 0.012$ rad/min, $\Omega_n = 1$ rad/min; \textbf{(v)}, $\Omega_v = 1$ rad/min, $\Omega_n = 0.001$ rad/min; \textbf{(vi)}, $\Omega_v = 1$ rad/min, $\Omega_n = 1$ rad/min, also shown in panel \textbf{c}. Cells are colored according to their orientation shown on the colorbar. For all simulations $\sigma_\textrm{iso}^2 = 2$ and $\sigma_\textrm{ani}^2 = 4$.}
\end{figure*}

\subsection{\label{subsec:nemaic_alignment_effect}Nematic alignment to neighbors improves ‘vertical’ alignment and directedness}

Following experimental motivation \cite{duclos2014perfectNemOrdInSpindleShapedFibroblasts, duclos2017fibroblastsOn+1defect, kaiyrbekov2023migration}, we have assumed that our cells' long axes have a nematic alignment controlled by $\Omega_n$. Can cells use this alignment to work together to align themselves perpendicular to the field, and increased directionality? %
We simulate cluster migration over a broad spectrum of alignment rates to nearest neighbors $\Omega_n$ and average velocity $\Omega_v$ in Fig. \ref{fig:phase_diagrams}. In addition to directionality $\langle \cos \theta \rangle$ and extent of vertical alignment of cells $\langle |\phi|\rangle$, we quantify the overall alignment within a group of cells using the nematic order parameter $Q$ \cite{duclos2014perfectNemOrdInSpindleShapedFibroblasts,duclos2017fibroblastsOn+1defect}
\begin{equation}
\label{eq:order_parameter}
   Q = \sqrt{\langle \cos 2 \phi \rangle^2 + \langle \sin 2 \phi \rangle^2 } 
\end{equation}
where $\phi$ represents the angle of cell orientation, and the averaging is performed across the cell population. $Q = 1$ means the long axes of the cells are perfectly aligned -- but not necessarily vertically aligned -- while $Q = 0$ if $\phi$ is uniformly distributed.

The cell directionality $\langle \cos \theta \rangle$ depends on both alignment to velocity $\Omega_v$ and alignment to neighbors $\Omega_n$ (Fig. \ref{fig:phase_diagrams}a). Directionality is maximal when alignment to velocity is in our intermediate range ($\Omega_v \sim 10^{-2}-10^{-1}$ rad/min) but alignment to neighbors is high ($\Omega_n \sim 10^{-1}-10^{0}$ rad/min). The directionality largely reflects the degree to which cells successfully reach a vertical orientation (Fig. \ref{fig:phase_diagrams}b). We know that in the absence of any alignment to velocity $\Omega_v \to 0$, cells are essentially randomly oriented relative to the field. Consistent with this idea, we see our lowest directionality and lowest $\langle |\phi |\rangle$ at low $\Omega_v \sim 10^{-3}$ rad/min. Increasing alignment to neighbors while keeping $\Omega_v$ low does make cells line up nematically (Fig. \ref{fig:phase_diagrams}c plots $Q$, and the difference is dramatic in comparing Fig. \ref{fig:phase_diagrams}i)and Fig. \ref{fig:phase_diagrams}ii). However, increasing $\Omega_n$ with low $\Omega_v$ fails to induce the vertical alignment  necessary to significantly improve directionality (Fig. \ref{fig:phase_diagrams}b).  

There are two key features in Fig. \ref{fig:phase_diagrams}a we want to mention. First, we see that the dependence of directionality on alignment to velocity $\Omega_v$ is not monotonic. At low $\Omega_v$, cells fail to orient relative to the field, and gain no significant directionality benefit, but at excessively high values of $\Omega_v$, directionality again decreases. We think this is because at high $\Omega_v$, cells rapidly align their long axis to be exactly perpendicular to their own averaged velocity, neglecting information from their neighbors. Secondly, we see that -- at sufficiently large $\Omega_v$ -- increasing alignment to neighbors $\Omega_n$ increases directionality. This suggests that cells are able to effectively share information by using their nematic alignment.

These key trends are somewhat robust to varying degrees of cell adhesiveness. Whether considering cells with weak adhesion ($k$ = 0.05 $\textrm{min}^{-1}$, Fig. \ref{fig:phase_diagrams_k005}) or those bound by stronger forces modeled as taut springs ($k$ = 1 $\textrm{min}^{-1}$, Fig. \ref{fig:phase_diagrams_k1}), the influence of $\Omega_v$ on vertical alignment is clear. In both scenarios, low $\Omega_v$ values fail to produce significant vertical alignment across a broad range of alignment rates to neighbors $\Omega_n$. However, for cells interconnected by taut springs, the non-monotonic trend is weaker -- rapid alignment at large $\Omega_v$ doesn't greatly impede directionality. We suspect this is  because in tightly-adherent clusters, cells have less noisy velocities -- they closely track the center of mass -- so cells benefit less from paying attention to their neighbors' orientations.

\begin{figure*}
\centering
\includegraphics[width=0.9\textwidth]{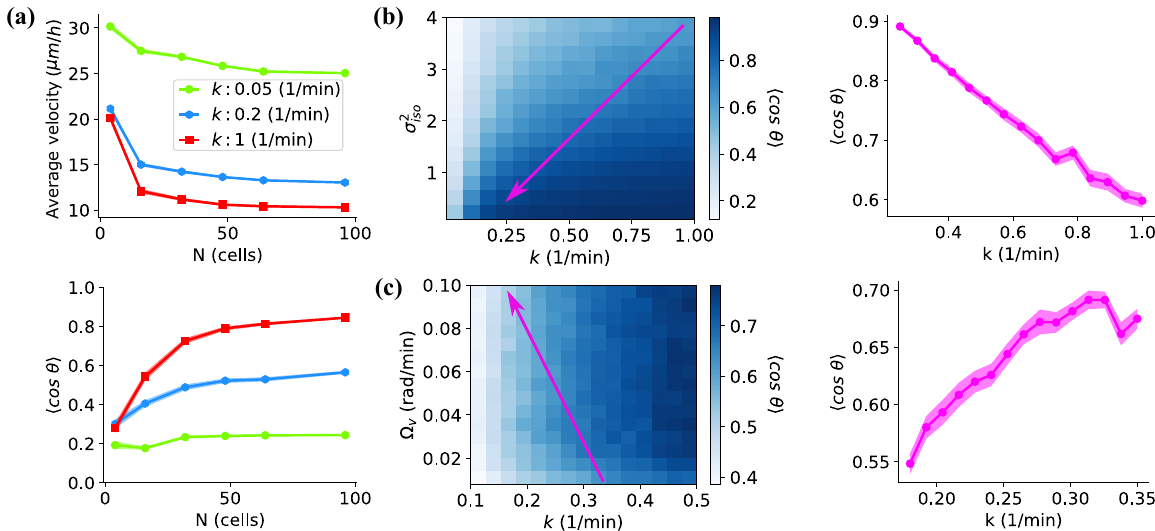}%
\caption{\label{fig:spring_dep} Dependence of directionality on cell-cell adhesion. \textbf{(a)} Directionality (top row) and velocity (bottom row) as a function of cell count for different (color coded) spring constants. \textbf{(b)} Directionality for different combinations of isotropic variance $\sigma_\textrm{iso}^2$ and interaction strength $k$ (left). The arrow indicates a linear path of concurrent variations of $\sigma_\textrm{iso}^2$ and $k$ resulting decreasing directionality shown on the right figure. Here $\Omega_v = 0.01 \ (rad/min)$.  \textbf{(c)} Directionality for different combinations of alignment rate to velocity $\Omega_{v}$ and interaction strength $k$ (left).  The arrow indicates a linear path of concurrent variations of $\Omega_v$ and $k$ resulting decreasing directionality shown on the right figure. Here $\sigma_\textrm{iso}^2 = 2$. 
The reported values represent averages across 40 simulations with 64 cells $\sigma_\textrm{ani}^2 = 4$. Where applicable the shaded areas represent standard errors of the mean. }
\end{figure*}

\subsection{Cell-cell adhesion is crucial for robust directedness of cells \label{sec:adhesion}}

We have noted earlier in this paper that cell-cell adhesion may play a role in controlling directionality of collective galvanotaxis. What do experiments suggest? %
Research from Min Zhao's group demonstrated that disrupting E-cadherin junctions in MDCK I cells with the monoclonal antibody DECMA-1 leads to increased cell speeds but diminished galvanotactic directionality \cite{li2012cadherin}. Similarly, keratinocyte speeds rise upon the disruption of cell adhesions by DECMA-1 \cite{shim2021overriding}. However, the impact on E-cadherin blockade by DECMA-1 on directionality varies significantly with the initial strength of cell-cell adhesion: disruption of E-cadherin junctions enhances the directionality of cells with strong adhesions (high calcium in media), has negligible effects on cells with medium adhesive interactions, and reduces directionality in cells with weaker adhesions (low calcium media). Can we understand why groups of cells might, depending on context, either have an increase or decrease in directionality as adhesion strengths are varied?

With our default parameters, cell speeds decrease and directionalities increase as adhesion strength is increased (Fig. \ref{fig:spring_dep}a). As we saw above, directionality typically increases with the number of cells in the group -- though at the weakest adhesion strength, where cells are often separate from the group, cell number only weakly influences directionality. Our cell speed results are consistent with experiment \cite{li2012cadherin,shim2021overriding}, while the directionality results are consistent with those of \cite{li2012cadherin} on MDCK I cells but do not explain the complicated dependence of directionality on adhesion seen in \cite{shim2021overriding}. Why not? One hypothesis is that changing cell-cell adhesion in the experiment regulates multiple factors at once -- so that we should model these experimental changes in cell-cell adhesion as changing multiple parameters in our model. For instance, changing cadherin expression is known to regulate cell shape in complex and context-dependent ways \cite{lawson2021jamming}. Similarly, changes in cell-cell adhesion may alter cell-substrate interactions \cite{burute2012spatial}. We explore these two hypotheses in our model, to see if it is possible to create decreases in directionality with increasing adhesion strength, or a non-monotonic trend.

We first study what happens if changing E-cadherin strength simultaneously changes cell size or shape, thus altering the precision of cells' ability to estimate the field direction \cite{nwogbaga2023physical}. As an example for what could happen if cell size changes as E-cadherin strength changes, we show how directionality varies if we vary both $\sigma^2_\textrm{iso}$ and $k$ in our model in Fig. \ref{fig:spring_dep}b, left. (This is only one possibility; models extending the work of \cite{nwogbaga2023physical} would be necessary to explicitly connect cell size and shape and accuracy.) Increasing the variance of the cells' estimates naturally leads to reduced directionality across a broad spectrum of adhesion strengths (see Fig. \ref{fig:spring_dep}b, left). As an illustration, suppose that when E-cadherin strength is regulated, this also changes $\sigma^2_\textrm{iso}$, following the pink arrow in Fig. \ref{fig:spring_dep}b, left. The directionalities that would be measured as E-cadherin is varied are then shown in Fig. \ref{fig:spring_dep}b, right. If we assume that longer keratinocytes have enhanced overall precision of electric field direction estimate (i.e. reduced $\sigma_\textrm{iso}^2$), then noting that keratinocytes become more elongated as adhesions weaken \cite{shim2021overriding}, this combination of effects could potentially explain the observed increase in directionality upon disrupting cell-cell adhesions.

A second hypothesis is that cell-cell and cell-substrate interactions may compete \cite{burute2012spatial}. Within our model, this might mean that cells with weaker cell-cell adhesion are more effective at aligning perpendicular to their own velocity. We plot the effect of varying $\Omega_v$ and $k$ separately in Fig. \ref{fig:spring_dep}c, left. Following the pink arrow in the phase diagram, which shows $\Omega_v$ increasing with decreasing $k$, leads to a weakly non-monotonic behavior, with a  slight dip in directionality at $k \sim 0.33$ $\textrm{min}^{-1}$ (Fig. \ref{fig:spring_dep}c, right). This non-monotonic behavior occurs because of the competition between the effect of the spring constant $k$, which increases accuracy and the decrease in $\Omega_v$, which reduces alignment perpendicular to the field and therefore accuracy. The decrease in directionality we see here is not dramatic, but is consistent with experimental findings of a minor dip in directionality under similar conditions.

\section{Discussion\label{sec:discussion}}

We began by hypothesizing that cells are better at sensing the orientation of an electric field when their long axes are orthogonal to the field. Given this assumption, we found that groups of cells can work together to improve their ability to sense fields by a combination of orienting perpendicular to their (time-averaged) velocity as well as aligning nematically with their neighbors, performing best when there is a balance between these two effects. Experiments have previously found inconsistent results on how galvanotactic directionality of cell groups depends on cell-cell adhesion \cite{shim2021overriding,li2012cadherin}. Our work proposes plausible mechanisms -- cell-cell adhesion regulating cell shape or cell-substrate interactions -- that could explain why, depending on context, cell-cell adhesion could either increase or decrease directionality.

While we have largely focused on the response of groups of cells to electric fields, our  assumptions are intentionally generic, and could be easily adapted to model other migration modes, including chemotaxis, haptotaxis, and durotaxis. The core assumption that would have to be updated is our assumption that cells only re-estimate the electric field direction every $\tau_\textrm{forget}$. This long timescale ($\sim 10$ min) is due to the time required to transport galvanotactic sensors along the cell surface \cite{allen2013electrophoresis,nwogbaga2023physical}. Chemotaxing cells, for instance, have a much shorter averaging time (estimated as a few seconds to tens of seconds for Dictyostelium \cite{fuller2010external,segota2013high,van2007biased}), and a more detailed model for cell polarization might have to be put into place. 

Our foundational assumption in our model, inspired by theoretical studies \cite{GeometryBiasGradientSensing}, is that cells display significant anisotropy in precision depending on their orientations. To some extent, this is unavoidable -- cells with larger length parallel to or perpendicular to the gradient will naturally have a larger ``signal'' of difference in electric potential or chemical concentration. However, experimental evidence supporting this assumption is limited, to our knowledge, though there are some intriguing hints that chemotaxis is cell shape-dependent \cite{morrow2019integrating}. Experiments directly measuring the effect of cell shape and orientation on chemotaxis or galvanotaxis may be difficult, because cells do not respond to an applied signal with a fixed shape -- they tend to elongate perpendicular to the signal. Separating these factors may require experimentalists to use micropatterning to constrain cell shapes \cite{thery2010micropatterning}; relatedly, dynamic micropatterns have recently been used to determine the extent to which cell shape can predict future migration direction \cite{isomursu2023dynamic}. 

In our current model, we have not integrated mechanisms by which cells' polarity can influence one another \cite{camley2017physical} like  contact inhibition of locomotion (CIL) \cite{stramer2017mechanismsOfCIL}, where cells typically repolarize in the opposite direction upon contact. Including CIL may allow for additional methods for how groups of galvanotaxing cells may improve their directionality, as previously studied for collective chemotaxis \cite{camley2018collective,camley2016emergent,ipina2022collective}. 
Additionally, our model of physical cell-cell interactions is relatively simple, with only a spring interaction for adhesion, and a phenomenological nematic alignment term. We also have not included cell deformability. These interactions could be treated by applying variants of the Gay-Berne potential \cite{cleaver1996extensionGayBerne} as in \cite{kaiyrbekov2023migration}, but at the cost of significantly increased complexity and number of parameters. %

\begin{acknowledgments}
The authors acknowledge support from NSF PHY 1915491. This work was carried out at the Advanced Research Computing at Hopkins (ARCH) core facility  (rockfish.jhu.edu), which is supported by the National Science Foundation (NSF) grant number OAC 1920103. We thank Emiliano Perez Ipi\~na and Ifunanya Nwogbaga for a close reading and comments on the draft.
\end{acknowledgments}

\appendix

\section{Details of numerical methods}

\subsection{Initialization of simulations}

We conduct $\mathcal{S} = 40$ simulations for each parameter configuration, with simulations indexed from 1 to 40. . To enhance comparability and reproducibility, we initialize the random seed to its corresponding simulation index at the beginning of each simulation. Cells are initially placed at random coordinates, maintaining a minimum distance of 25 $\mu m$ from each other, with a density of $\rho_i = 1024 \ cells/mm^2$ near the center of the simulation box. This setup guarantees no isolated cells and fosters the formation of a single cell cluster at the start. The initial orientations of the cells are randomly assigned, following a uniform distribution  $\mathcal{U} \left[ 0, 2 \pi \right]$.
\subsection{Numerical integration}

In order to monitor a cell's neighboring cells, we maintain a list containing all cells within a distance  $r_{nn}$ of the given cell.  This list is regularly updated each time any cell moves a distance of $\Delta d_{nn}$ since the previous update. It's important to note that a cell doesn't interact with all the cells on the neighbor list, but only with those within the defined cutoff distance $r_c$. 

{At the beginning of each iteration, we calculate the interaction forces and torques acting on every cell. Subsequently, we employ numerical integration to solve the equations of motion as follows: For each cell $i$, the positions are updated every simulation step as:}
\begin{equation}
 {\bf r}^i (t + \Delta t) = {\bf r}^i (t) + \left( {\bf p}^i +  \sum_{j \underset{n}{\sim} i} {\bf F}^{ij} \right) \Delta t
\label{eq:eqofmotion_code}
\end{equation}
where as before the sum is over the cells that are within the interaction cutoff distance $r_{c}$. {Cells synchronously update their polarity at intervals of $\tau_\textrm{forget}$, keeping this updated polarity constant for the subsequent period of $\tau_\textrm{forget}$.} Cell orientations are updated via Euler-Maruyama method \cite{kloeden1992stochasticDiffEQN}:
\begin{eqnarray}
\nonumber
\phi^i (t + \Delta t) =  && \phi^i (t)   
                        \\&&- \ \Omega_{v} \sin(2[\phi^i - (\alpha_{\langle v^i \rangle_T} + \pi/2)]) \Delta t \nonumber\\&& - \ \Omega_n \sum_{j \underset{n}{\sim} i} \sin(2[\phi^i - \phi^j]) \Delta t 
                        \nonumber\\&& 
                        + \sqrt{2 D_r} \Gamma
\label{eq:eqofrotation_code}
\end{eqnarray}
where the summation is over neighbor cells. {In each iteration, cells update $\alpha_{\langle v^i \rangle_T}$ to retain a moving average of the angle.} 
The term $\Gamma = \int_t^{t+\Delta t} \xi_r(t) dt$, hence $\langle \Gamma \rangle = 0$  and  $\langle \Gamma^2 \rangle = \Delta t$. Numerically, every simulation timestep we sample $\Gamma$ from normal distribution $\mathcal{N}(\mu = 0, \sigma^2 = \Delta t)$ i.e.  $\Gamma \sim \mathcal{N}(\mu = 0, \sigma^2 = \Delta t)$.

\subsection{Parameter setting}

{We think of our cells as ellipses having a long axis radius of $a$ = 20 $\mu m$ and short axis radius of $b$ = 10 $\mu m$. We set the interaction cutoff distance ($r_c$), spring equilibrium length ($r_{eq}$), and neighbor list tracking radius ($r_{nn}$) were set to ensure most interactions are only with reasonable neighboring cells, given the cell size we defined. However, $a$ and $b$ do not actually affect the simulation, only the plotting. Within the simulation, the cell is only characterized by its long axis orientation $\phi$, and the cell-cell interactions are simple isotropic springs, which set the cell-cell distances.} The rotational diffusion coefficient was calibrated to ensure cells exhibited behavior consistent with typical observations. We performed simulations for various $D_r$ values and interaction strengths using our default choice of variances. Consistent with expectations cells were more aligned for smaller diffusion coefficient (Fig. \ref{fig:diff_dep}a,d,g) and disordered for large $D_r$ (Fig. \ref{fig:diff_dep}c,f,i).  We select value of $0.003 \ rad^2/min$ (see Fig. \ref{fig:diff_dep}b,e,f) since its orientation distributions for medium interaction strength $k = 0.2 \ \textrm{min}^{-1}$ roughly resembled experimental histograms \cite{zajdel2020scheepdog}. The complete set of parameters is detailed in Table \ref{tab:parameters}. Parameters not listed in this table were varied for the purposes of this study, and their specific values are provided in the main text.

\begin{table*}
\caption{\label{tab:parameters}Simulation parameters}
\begin{ruledtabular}
\begin{tabular}{lll}
 Parameter&Meaning&Value\\ \hline

 $r_{c}$&Interaction cutoff radius& 50 $\mu m$  \\
 $r_{eq}$& Equilibrium length of spring &40 $\mu m$ \\
 $r_{nn}$&Cutoff radius of neighbor lists& 60 $\mu m$ \\
 $\Delta d_{nn}$& Threshold displacement to initiate neighbor list update & 8 $\mu m$\\
 $D_r$&Rotational diffusion coefficient& 0.003 $rad^2 / min$\\
 $\tau_\textrm{forget}$& Cell repolarization timescale (period) & 10 $min$ \\
 $\Delta t$&Simulation time step & 0.001 $min$ \\
 
\end{tabular}
\end{ruledtabular}
\end{table*}

\providecommand{\noopsort}[1]{}\providecommand{\singleletter}[1]{#1}%

\clearpage
\appendix
\onecolumngrid
\renewcommand{\thefigure}{S\arabic{figure}}
\setcounter{figure}{0}

\section{Center of mass motion \label{app:cm}}
The center of mass (c.o.m.) of a cell cluster evolves via simple equation of motion, which can be computed by taking sum of Eqn. (\ref{eq:eqofmotion}) over cells:
\begin{equation}
\label{eq:COM_eqofmotion_computation}
     \frac{1}{N} \sum_{i=1}^N \frac{\partial {\bf r^i}}{\partial t} = \frac{1}{N} \sum_{i=1}^N {\bf p^i} + \frac{1}{N} \sum_{i=1}^N \sum_j {\bf F}^{ij}
\end{equation}
Owing to Newton's third law, the last term cancels out, i.e., $\sum_{i=1}^N \sum_j {\bf F}^{ij}=0$. 
Incorporating summation inside the partial derivative of Eqn. (\ref{eq:COM_eqofmotion_computation}) and defining the coordinate of the center of mass  as $\mathbf{r}^{c.o.m} = \sum_{i=1}^N \mathbf{r}^i / N$ we reach:
\begin{equation}
\label{eq:COM_eqofmotion}
      \frac{\partial \mathbf{r}^{c.o.m}}{\partial t} = \frac{1}{N} \sum_{i=1}^N \mathbf{p}^i
\end{equation}
Since a cell $i$ polarizes towards its estimate of field direction (i.e. ${\bf p}^i  = \cos (\zeta^i) \hat{x}  +  \sin( \zeta^i) \hat{y}$, where $\zeta^i \sim VM(0, \kappa^i)$), the center of mass migrates towards groups' the mean estimated direction. Due to law of large numbers, the accuracy of average estimate improves for larger $N$, hence the c.o.m. of larger groups of cells follow electric field more efficiently.

The migration of the c.o.m. is independent of the adhesion strength $k$; however, $k$ influences the extent to which individual cells adhere to the c.o.m. within the displacement measurement interval 0f $30$ minutes. A higher $k$ ensures that cells robustly track the c.o.m., whereas a lower $k$ allows individual cells more freedom to deviate from the group's migration path.

\begin{figure*}
\centering
\includegraphics[width=0.95\textwidth]{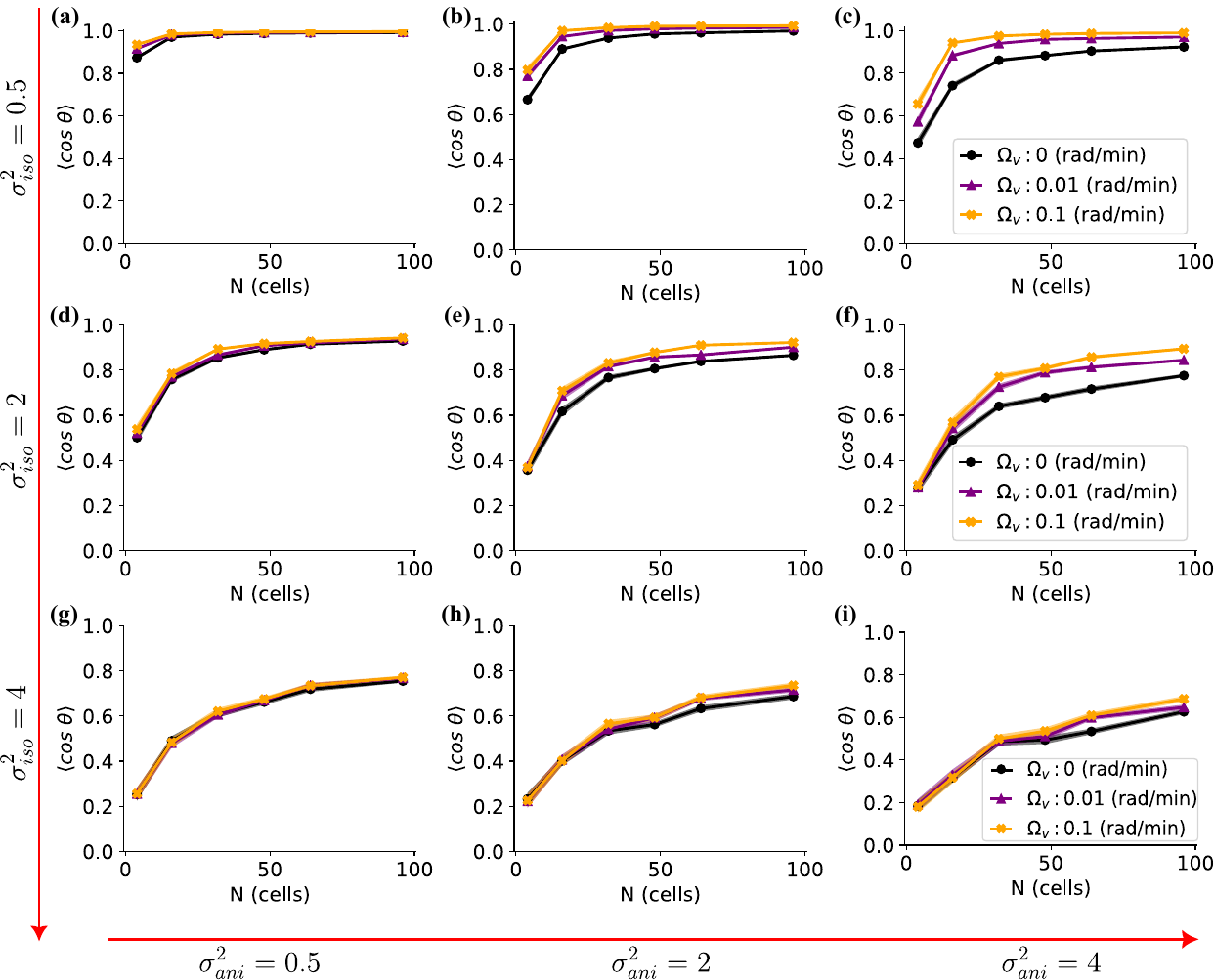}%
\caption{\label{fig:variance_dep_k1}  The average directionality $  \langle \cos\theta \rangle$ for various combinations of isotropic $\sigma_\textrm{iso}^2$ and anisotropic $\sigma_\textrm{ani}^2$ variances and alignment rates to velocity. This plot is the same as Fig. 1 but at cell-cell the interaction strength of $k$ = 1 $\textrm{min}^{-1}$.  
Isotropic component ($\sigma_\textrm{iso}^2$) changes across  rows (top  to bottom) and anisotropic component ($\sigma_\textrm{ani}^2$) across rows (left to  right) with  specific values shown at right side and bottom of the figure (i.e. Figure (f) show directionalities for $\sigma_\textrm{iso}^2 = 2, \sigma_\textrm{ani}^2 = 4$ ). The averages are over 40  simulations and each simulation is performed with 64 cells. For each simulation the reported directionality is the steady state average over final 5 hours  of simulation Fig. \ref{fig:intro_fig}f. Results for different vaues of alignment rates to average velocity are color coded $\Omega_v = 0$ rad/min (black), $\Omega_v = 0.01$ rad/min (purple), and $\Omega_v = 0.1$ rad/min (orange). The averaging time $T$ for velocity is set to 1h.  The shaded areas represent standard errors, although they may not be easily discernible due to their small size. }
\end{figure*}

\begin{figure*}
\centering
\includegraphics[width=0.95\textwidth]{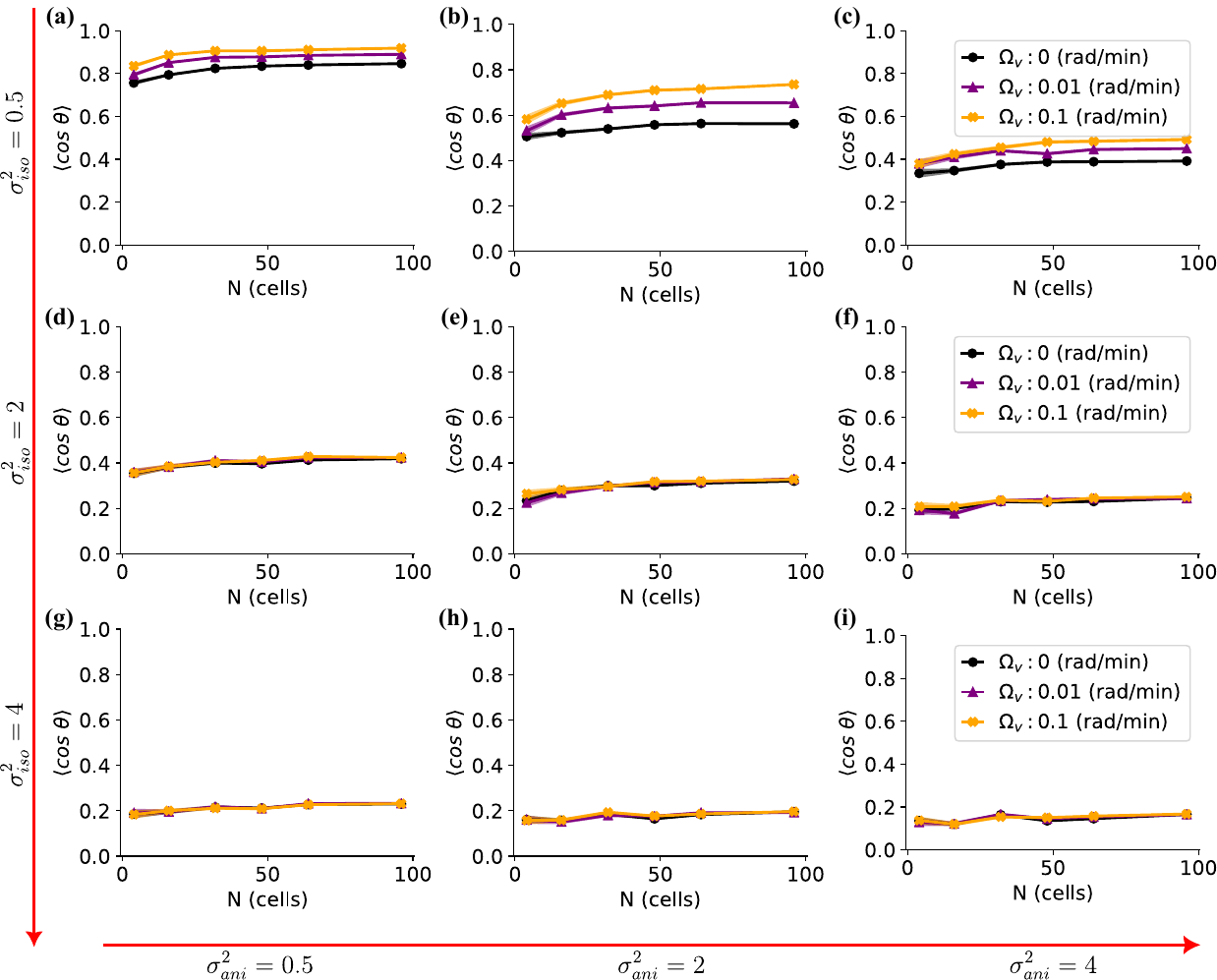}%
\caption{\label{fig:variance_dep_k0.05}The average directionality $  \langle \cos\theta \rangle$ for various combinations of isotropic $\sigma_\textrm{iso}^2$ and anisotropic $\sigma_\textrm{ani}^2$ variances and alignment rates to velocity. This plot is the same as Fig. 1 but with a cell-cell interaction strength of $k$ = 0.05 $\textrm{min}^{-1}$.   
Isotropic component ($\sigma_\textrm{iso}^2$) changes across  rows (top  to bottom) and anisotropic component ($\sigma_\textrm{ani}^2$) across rows (left to  right) with  specific values shown at right side and bottom of the figure (i.e. Figure (f) show directionalities for $\sigma_\textrm{iso}^2 = 2, \sigma_\textrm{ani}^2 = 4$ ). The averages are over 40  simulations and each simulation is performed with 64 cells. For each simulation the reported directionality is the steady state average over final 5 hours  of simulation Fig. \ref{fig:intro_fig}f. Results for different vaues of alignment rates to average velocity are color coded $\Omega_v = 0$ rad/min (black), $\Omega_v = 0.01$ rad/min (purple), and $\Omega_v = 0.1$ rad/min (orange). The averaging time $T$ for velocity is set to 1h.  The shaded areas represent standard errors, although they may not be easily discernible due to their small size. }
\end{figure*}

\begin{figure}
\includegraphics[width=0.70\textwidth]{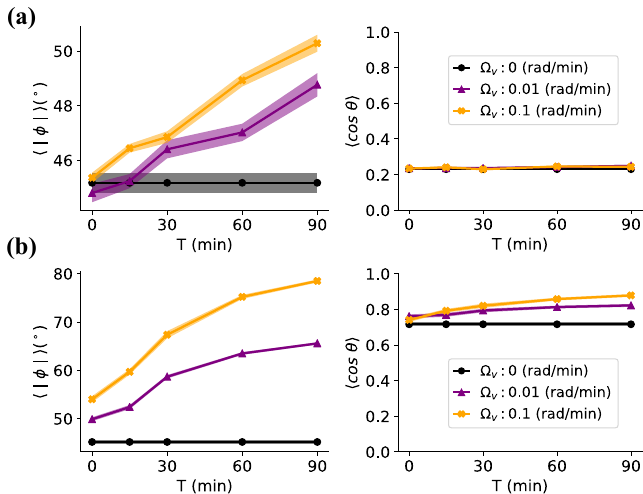}%
\caption{\label{fig:mem_dep_supp} Cell alignments and directionalities as a function of velocity averaging time. The reported values represent averages across 40 simulations of 64 cells with $\sigma_\textrm{iso}^2 = 2 \ \& \ \sigma_\textrm{ani}^2 = 4$ conducted at an interaction strength of \textbf{(a)} $k$ = 0.05 $\textrm{min}^{-1}$ and \textbf{(b)} $k$ = 1 $\textrm{min}^{-1}$. In the left column, the absolute value of the cell alignment angle is presented, while the right column displays the corresponding directionality.  The shaded areas represent standard errors of the mean.}
\end{figure}

\begin{figure*}
\centering
\includegraphics[width=0.95\textwidth]{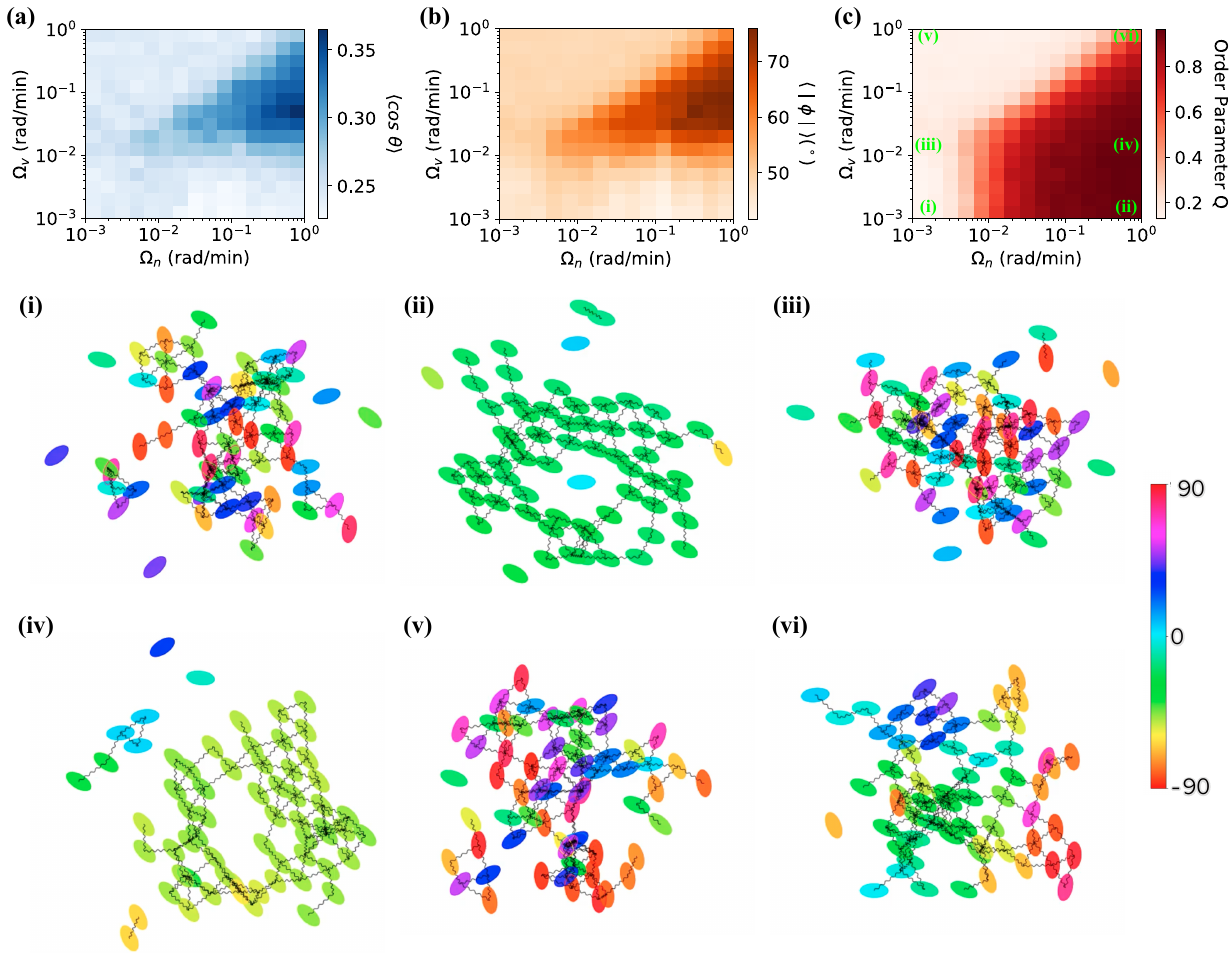}%
\caption{\label{fig:phase_diagrams_k005}Effect of rates of alignment to velocity ($\Omega_v$) and neighbors ($\Omega_n$) on directionality \textbf{(a)}, alignment \textbf{(b)} and order parameter \textbf{(c)}. Each grid value represents an average result over 40 simulations conducted with 64 cells at the interaction strength of k = 0.05 $\textrm{min}^{-1}$ with an averaging time 
$T$ = 1 h with colorbars indicating corresponding numeric values. Example simulation snapshots for alignment rate tuples of \textbf{(i)}, $\Omega_v = 0.001$ rad/min, $\Omega_n = 0.001$ rad/min; \textbf{(ii)}, $\Omega_v = 0.001$ rad/min, $\Omega_n = 1$ rad/min; \textbf{(iii)}, $\Omega_v = 0.012$ rad/min, $\Omega_n = 0.001$ rad/min; \textbf{(iv)}, $\Omega_v = 0.012$ rad/min, $\Omega_n = 1$ rad/min; \textbf{(v)}, $\Omega_v = 1$ rad/min, $\Omega_n = 0.001$ rad/min; \textbf{(vi)}, $\Omega_v = 1$ rad/min, $\Omega_n = 1$ rad/min, also shown in panel \textbf{c}. Cells are colored according to their orientation shown on the colorbar. For all simulations $\sigma_\textrm{iso}^2 = 2$ and $\sigma_\textrm{ani}^2 = 4$.}
\end{figure*}

\begin{figure*}
\centering
\includegraphics[width=0.95\textwidth]{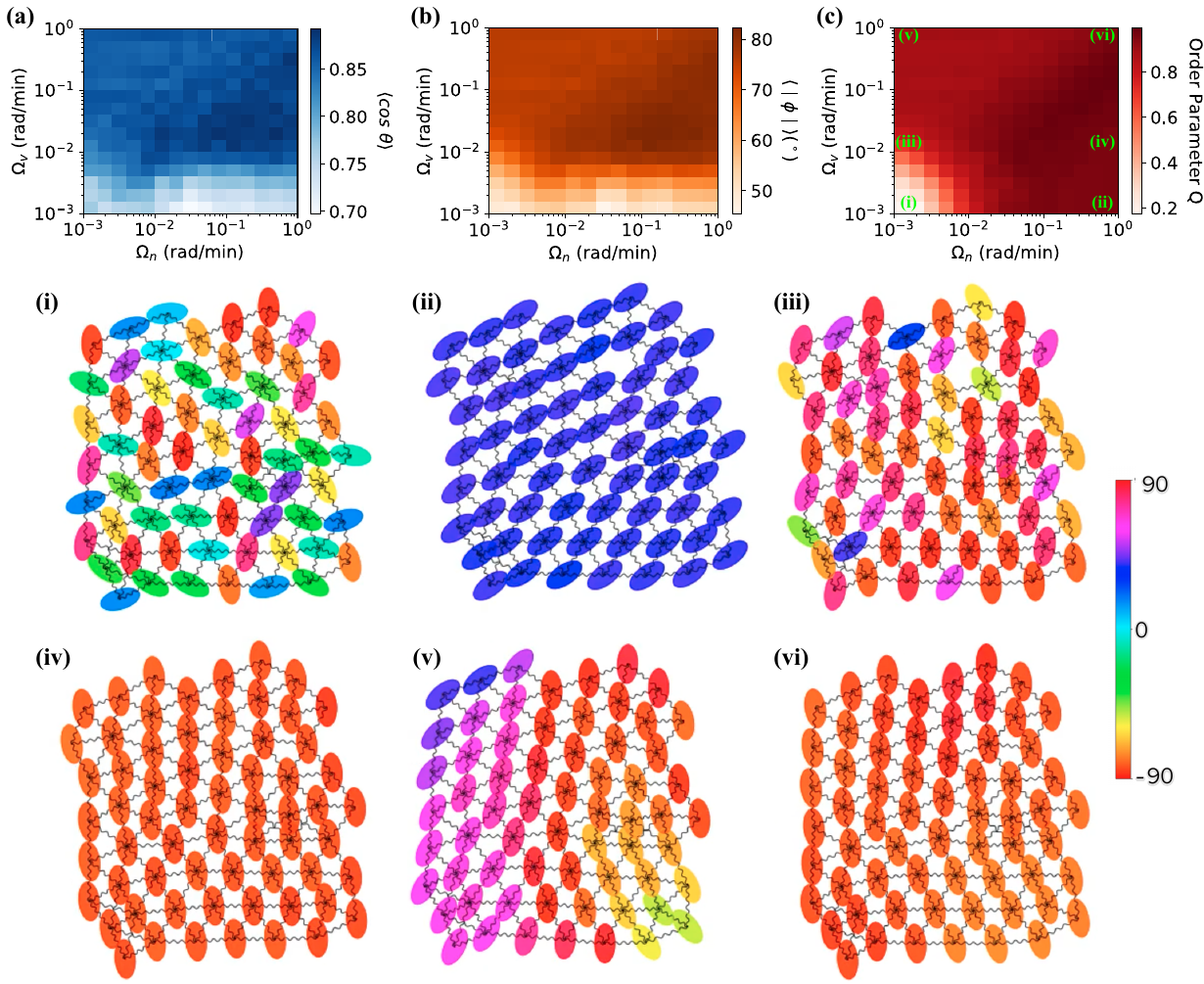}%
\caption{\label{fig:phase_diagrams_k1}Effect of rates of alignment to velocity ($\Omega_v$) and neighbors ($\Omega_n$) to directionality \textbf{(a)}, alignment \textbf{(b)} and order parameter \textbf{(c)}. Each grid value represents an average result over 40 simulations conducted with 64 cells at the interaction strength of k = 1 $\textrm{min}^{-1}$ with an averaging time 
$T$ = 1 h with colorbars indicating corresponding numeric values. Example simulation snapshots for alignment rate tuples of \textbf{(i)}, $\Omega_v = 0.001$ rad/min, $\Omega_n = 0.001$ rad/min; \textbf{(ii)}, $\Omega_v = 0.001$ rad/min, $\Omega_n = 1$ rad/min; \textbf{(iii)}, $\Omega_v = 0.012$ rad/min, $\Omega_n = 0.001$ rad/min; \textbf{(iv)}, $\Omega_v = 0.012$ rad/min, $\Omega_n = 1$ rad/min; \textbf{(v)}, $\Omega_v = 1$ rad/min, $\Omega_n = 0.001$ rad/min; \textbf{(vi)}, $\Omega_v = 1$ rad/min, $\Omega_n = 1$ rad/min, also shown in panel \textbf{c}. Cells are colored according to their orientation shown on the colorbar. For all simulations $\sigma_\textrm{iso}^2 = 2$ and $\sigma_\textrm{ani}^2 = 4$.}
\end{figure*}

\begin{figure*}
\centering
\includegraphics[width=0.95\textwidth]{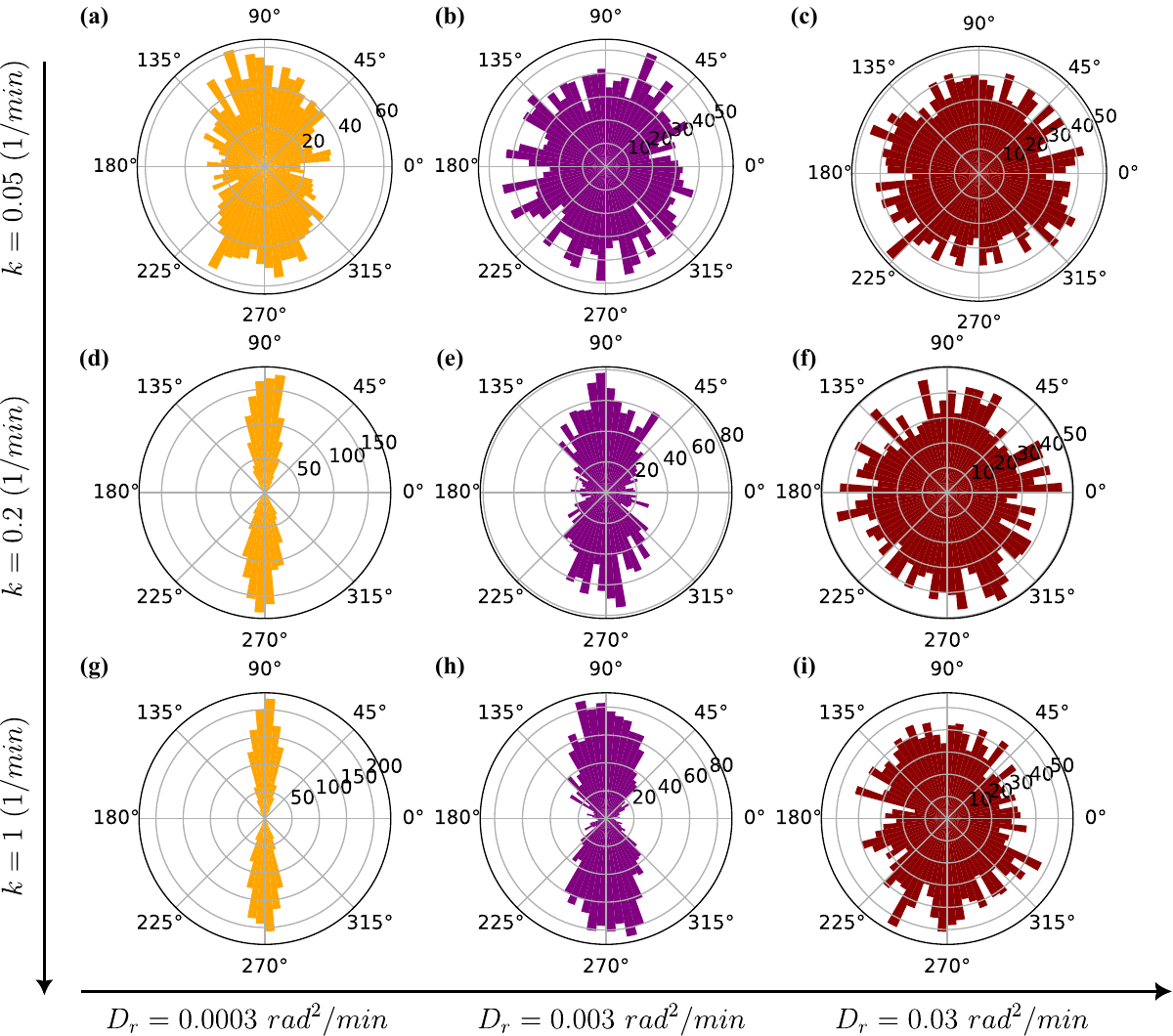}%
\caption{\label{fig:diff_dep}
Cell orientation histograms for various combinations of interaction strengths and diffusion coefficients. interaction strength varies across  rows (top  to bottom) and diffusion coefficient across columns (left to  right) with  specific values of spring constant shown at right side and diffusion coefficient at bottom of the figure (i.e. panel \textbf{(e)} shows orientations for $k = 0.2 \ \textrm{min}^{-1}$ and $D_{r} = 0.003 \ rad^2/min$ ). {Each histogram is compiled from the final snapshot data of 40 simulations, each featuring 64 cells.}  Cells do not align to neighbors $\Omega_n = 0 \ rad/min$, align to average velocity (averaging time $T$ = 1 h) at the rate of $\Omega_v = 0.01 \ rad/min$, and have $\sigma_\textrm{iso}^2 = 2$ and $\sigma_\textrm{ani}^2 = 4$.
}
\end{figure*}

\end{document}